\documentclass[prb,preprint,eqsecnum]{revtex4}
\usepackage{amsmath}
\usepackage{graphicx}
\usepackage{amssymb}

\usepackage{graphicx}
\usepackage{amsfonts}%

\begin{document}

\title{Effects of columnar disorder on flux-lattice melting in high-temperature
superconductors}

\author{Sandeep Tyagi}
\author{Yadin Y. Goldschmidt}
\affiliation{Department of Physics and Astronomy, University of Pittsburgh,
Pittsburgh, Pennsylvania 15260}


\begin{abstract}

The effect of columnar pins on the flux-lines melting transition in
high-temperature superconductors is studied using Path Integral 
Monte Carlo simulations. We highlight the similarities and differences in the effects
of columnar disorder on the melting transition in YBa$_2$Cu$_3$O$_{7-\delta}$
(YBCO) and the highly anisotropic Bi$_2$Sr$_2$CaCu$_2$O$_{8+\delta}$ (BSCCO) 
at magnetic fields such that the mean separation between flux-lines is
smaller than the penetration length. For pure systems, a 
first order transition from a flux-line solid to a liquid phase is seen 
as the temperature is increased.
When adding columnar defects to the system, the transition temperature
is not affected in both materials as long as the strength of
an individual columnar defect (expressed as a flux-line defect interaction) 
is less than a certain threshold for a given density of randomly distributed
columnar pins. This threshold strength is lower for YBCO than for
BSCCO. For higher strengths the transition line is shifted for both 
materials towards higher temperatures, and the
sharp jump in energy, characteristic of a first order transition,
gives way to a smoother and gradual rise of the energy, characteristic
of a second order transition. Also, when columnar defects are present, the vortex 
solid phase is replaced by a pinned Bose glass phase and this is manifested by
a marked decrease in translational order and orientational order as measured by the 
appropriate structure factors. For BSCCO, we report an unusual rise
of the translational order and the hexatic order
just before the melting transition. No such rise is observed in YBCO.

\end{abstract}

\maketitle

\section{Introduction}
Type II superconductors \cite{tinkham,blatter,brandt} allow for a partial penetration 
of magnetic field into the bulk of the superconducting (SC) material when the 
applied field $H$ satisfies $H_{c1}<H<H_{c2}$. In a seminal work Abrikosov 
\cite{abrikosov} showed that when the ratio $\lambda/\xi$, where $\lambda$ 
is the magnetic field penetration depth and $\xi$ is the coherence length, 
is greater than 
$1/\sqrt{2}$ the magnetic field penetrates the SC material in the form of 
flux-lines (FLs). These FLs are also called vortices, since they are surrounded 
by circular currents. Each FL carries a quantized unit of flux $\phi_0=hc/2e$ 
called the fluxoid. The FLs have cylindrical shape of radius 
$\approx \lambda$ (the radius is not sharp since the magnetic field decays 
exponentially like $\exp(-r/\lambda)$, where r is the distance from the axis) 
and a non-SC core of radius $\approx \xi$. Due to a repulsive interaction 
among the Fls, they arrange themselves in a triangular lattice referred to 
as the vortex solid (VS). This result follows from mean-field theory. 

After high-temperature superconductors were discovered in the 1980's, it 
became apparent that thermal fluctuations, not included in the mean-field 
theory \cite{brezin}, play an important role at relatively high temperatures and fields, 
still below $T_c$ and 
$H_{c2}$. These fluctuations can cause the Abrikosov lattice to melt into 
a disordered liquid via a first order transition (FOT), which can be 
roughly estimated using the Lindemann criterion known from solid state 
physics \cite{lindemann,nelson3,pelcovits}. Technologically, the melting of the FL lattice 
is important since the vortex liquid (VL) is not actually SC due to the 
dissipation caused by the FL motion when an electric current passes through 
the system. Pinning of FLs by naturally occurring defects in the form of 
vacancies, interstitials, twin and grain boundaries etc., is effective to 
impede FL motion in the VS phase, where the FLs form a rigid correlated network. 
The effectiveness of the pinning manifests itself by leading to high critical 
currents. In the VL phase pinning of a few vortices does not inhibit others 
from moving when a current is applied. Thus for practical purposes the sudden 
increase in resistivity occurs at the melting transition rather than when 
$H=H_{c2}(T)$ for any reasonably non-vanishing currents. 

The existence of the melting transition in high-$T_c$ pristine materials has been 
established through numerous experimental \cite{exp1,exp2,cubitt,exp3,exp4} and 
numerical \cite{numer1,margo,ryu,koshelev,hu,nordborg,chin} studies. As was mentioned above, disorder in the 
form of points defects and sometimes more extended defects can and does occur 
naturally in laboratory samples. In addition artificial point defects can be 
induced by bombarding the sample with electrons originating from particle 
accelerators. Extended columnar defects in the form of linear damaged tracks 
piercing through the sample can be induced by heavy ion irradiation. Both 
naturally occurring and artificially induced defects are situated at random 
positions in the sample and their effective pinning strength (i.e. their 
interaction with FLs) can also vary from defect to defect. Thus defects 
play the role of quenched disorder. The adjective ``quenched'' refers to 
the immobility of these defects during experimental time scales.
Introduction of disorder in terms of point defects or columnar pins 
affects both the properties of the solid and liquid phases and might 
also shift the location of the melting transition in the H-T plane 
\cite{blatter,brandt}. In the case of point pins, the VS phase is replaced 
with a Bragg glass phase \cite{giamarchi,nattermann}, characterized by 
quasi-long-range order. The melting transition is predicted to shift
towards lower temperatures \cite{MD,ertas,goldschmidt}. In the case of
columnar pins the VS phase is replaced with a so called pinned Bose
glass \cite{boseglass} where FLs are trapped by the
columnar defects and the whole lattice becomes immobile. The Bose glass phase 
is similar to the localized phase of a two dimensional repulsive Bose gas in the
presence of quenched disorder, as will be explained in more detail in the next section. 

The effect of both kinds of disorders on the FLs melting has been studied 
experimentally in various high-temperature superconductors.
Two common materials that have been extensively investigated are YBa$_2$Cu$_3$O$_{7-\delta}$
(YBCO) and Bi$_2$Sr$_2$CaCu$_2$O$_{8+\delta}$ (BSCCO), 
both having critical temperatures ranging between 90-120 K at $H=0$. The main 
difference between these materials is their anisotropy parameter 
$\gamma^{2}=m_{z}/m_{\bot}>1$, where $m_{z}$ and $m_{\bot}$ denote the effective masses of 
electrons moving along the $c$-axis and the $ab$-plane respectively.  
BSCCO is much more anisotropic: its anisotropy parameter $\gamma$ lies in the
range of 50-200 compared to the range of 5-7 for YBCO \cite{blatter}. This 
fact causes the FLs to be much ``softer'' or elastic. Thus in the case 
of BSCCO the FLs are sometimes described as a collection of loosely connected 
``pancakes'' residing in adjacent Cu-O planes.
Experimental studies on YBCO have shown a marked shift in the irreversibility
line in the presence of the columnar disorder \cite{konczykowski,civale,samoilov,paulius,olsson}.
The irreversibility line in the H-T plane marks the onset of hysteresis effects and is located 
close to the melting transition on the solid phase side.
For BSCCO, many experimental studies have been conducted 
\cite{gerhauser,klein,konczykowski1,seow,khaykovich,lee,beek1,beek2}.
The more recent ones have shown \cite{khaykovich,lee,beek1,beek2} that the melting line is not 
shifted when the density of columnar defects is relatively low, $B_{\phi }<<B$,
but for $B_\phi \approx B$ a shift in the position of the melting
transition is observed. Here the matching field $B_{\phi }$ is defined as 
$B_{\phi }=n_{d}\phi_{0}$ 
where $n_{d}$ is the density of the columnar defects and $\phi_{0}$ is the 
flux quantum. 

Theoretical work on columnar disorder includes Bose Glass theory
\cite{boseglass}. Radzihovsky \cite{radzihovsky}
considered the possibility of two kinds of Bose glass phases (strongly
or weakly pinned) depending on whether $B < B_{\phi}$ or 
$B > B_{\phi }$. More recently, columnar as well as point disorder
were investigated by Goldschmidt \cite{goldschmidt} using replica field 
theory. He showed that the melting line shifts to lower temperature
in the case of point disorder and to higher temperature in the case of columnar disorder.

Due to the complexity of the problem, especially in the presence of disorder, simulations
have been very useful in studying the FL system.
There have been many simulation studies of the vortex system in the presence
of disorder. However, most simulation work has been confined to the
addition of point disorder only. In particular, there is little
work done on the effects of columnar disorder on the FL melting.
Recent work by Wengel and T\"auber \cite{wengel} concentrated on the case of high defect 
density region $B_{\phi}\approx B$ . In contrast, a recent
simulation study by Sen {\it et al.} \cite{sen} uses a small density of columnar 
defects, each of infinite strength, but considers an extremely small magnetic field. 
Similarly, Nandgaonkar {\it et al.} \cite{nandgaonkar} also investigate the case of a very 
small magnetic field.
In this paper we consider columnar defects of a finite strength ($\eta$) 
with a relatively low defect density, $B_{\phi}/B=0.2$, and with realistic magnetic fields 
as used in the experiments.

\section{The model}

We first discuss the method implemented for YBCO: 

Following Nelson \cite{nelson}, we map the system of $N$ vortices (FLs)
in a high-$T_c$ superconductor onto $N$ interacting bosons in 2-space + 1-time
dimensions. Then we do a Path Integral Monte Carlo (PIMC) simulation on this
system. The partition function of $N$ vortices
can be expressed as:
\begin{equation}
\Xi (T,L_{z},N)=e^{-F(T,L_{z},N)/kT}=\frac{1}{N!}\int 
\prod _{i=1}^{N}\emph {D}{\mathbf{R}_{i}(z)}e^{-\Im /kT}
\end{equation}
 where $\mathbf{R}_{i}(z)$ denotes the 2D position vector of the
i'th vortex at a height $z$ along the $c$-axis, $F$ is the free
energy as a function of temperature $T$, $L_{z}$ is the length of
the sample along the $z$-direction. The London free-energy functional
$\Im $ is given by
\begin{equation}
\frac{\Im }{kT}=\frac{1}{kT}\int _{0}^{L_{z}}dz\left\{ \sum _{i}
\frac{\varepsilon_{l}}{2}\left(\frac{d\mathbf{R}_{i}}{dz}\right)^{2}
+\sum _{i<j}2\varepsilon _{0}K_{0}\left(\frac{R_{ij}}{\lambda }\right)\right\} ,
\end{equation}
where $\varepsilon _{0}=\phi _{0}^{2}/$ $(4\pi \lambda )^{2}$ is
the vortex line energy per unit length, the line tension is $\varepsilon _{l}
=\varepsilon _{0}\ln ({a_{0}/2\sqrt{\pi }\xi )/\gamma }^{2}$
and $R_{ij}(z)=|\mathbf{R}_{i}(z)-\mathbf{R}_{j}(z)|$. Here
\begin{equation}
a_{0}=\sqrt{\frac{2\phi _{0}}{\sqrt{3}B}}
\end{equation}
is the lattice spacing, $B$ is the magnetic field along the $z$-direction and
$\gamma$ is the anisotropy parameter. $K_{0}(R_{ij}/\lambda )$ is the in-plane
interaction potential between two FLs a distance $R_{ij}$ apart. $K_{0}$ denotes
the modified Bessel function of the first kind. This expression for
the London free-energy is an approximation that neglects the
non-local interaction of real vortices and replaces it by in-plane
interactions only, which is really justified if the FLs do
not deviate too much from straight lines along the $z$-axis \cite{nordborg}.
With this approximation, the system of $N$ interacting FLs is equivalent to a 
system of $N$ bosons in $d=2$ dimension interacting
with a pairwise potential $K_{0}(R_{ij}/\lambda )$. 

The path integral representation of a system of $N$ bosons of mass
$m$ each in two dimensions, interacting through a potential $V(r)=g^{2}K_{0}(r/\lambda )$,
with $g$ being the strength and $\lambda$ being the range of the
repulsive interaction, is given at finite temperature $T_{B}$ in terms
of the imaginary time action
\begin{equation}
\frac{S}{\hbar }=\frac{1}{\hbar }\int _{0}^{\hbar /kT_{B}}
d\tau \left\{ \sum _{i}\frac{m}{2}\left(\frac{d\mathbf{R}_{i}}
{d\tau }\right)^{2}+\sum _{i<j}g^{2}K_{0}\left(\frac{R_{ij}(\tau )}
{\lambda }\right)\right\} .
\end{equation}
Here $\tau $ is the imaginary time and $T_{B}$ is the temperature
of the Bose system. We see that there is a one to one parameter mapping
between the boson system and the vortex system \cite{nordborg}:
\[
\tau \rightarrow z, \, \, \hbar \rightarrow kT, \, \, g^{2}\rightarrow 
2\varepsilon_{0},\, \, \hbar /kT_{B}\rightarrow L_{z},\, \, 
m \rightarrow \varepsilon _{l},\, \, n=N/A=2/(\sqrt{3}a_{0}^{2})=B/\phi _{0},
\]
where $n$ is the average density of bosons (and FLs) and $A$
is the area of the sample. We can write the London free-energy functional
in a dimensionless form as follows: 
\begin{equation}
\frac{\Im }{kT}=\int _{0}^{\beta }d\tau \left\{ \sum _{i}\frac{1}{2\Lambda^{2}}
\left(\frac{d\mathbf{R}_{i}}{d\tau }\right)^{2}+\sum _{i<j}K_{0}
\left(\frac{R_{ij}}{\tilde{\lambda }}\right)\right\} ,
\end{equation}
where 
\begin{equation}
\Lambda =\frac{\hbar }{a_{0}g\sqrt{m}}=\frac{kT}{a_{0}
\sqrt{2\varepsilon _{l}\varepsilon _{0}}}, \ \ \ \ 
\beta =\frac{g^{2}}{kT_{B}}=\frac{2\varepsilon _{0}L_{z}}{kT},
\ \ \ \ \tilde{\lambda }=\frac{\lambda }{a_{0}}.
\end{equation}
All lengths are measured in units of $a_{0}$ and energies in units
of $g^{2}$ for bosons and $\varepsilon _{0}a_{0}$ for FLs.
We next discretize the integral along the $z$-axis by dividing it into $M$
segments: 
\begin{equation}
\Xi (\beta ,\Lambda ,T)=\frac{1}{N!}\left(\frac{1}{2\pi 
\Lambda^{2}\tau }\right)\sum _{P}\int \prod _{m=1}^{M}
\prod _{i=1}^{N}d^{2}\mathbf{R}_{i,m}
e^{-\Im \lbrack {\mathbf{R}_{i,m}}]/kT}\label{Xi}
\end{equation}
where $\tau =\beta /M$, $m$ labels the planes and 
\begin{equation}
\frac{\Im \lbrack {\mathbf{R}_{i,m}}]}{kT}=\sum _{i,m}
\frac{\left(\mathbf{R}_{i,m+1}-\mathbf{R}_{i,m}\right)^{2}}
{2\Lambda^{2}\tau }+\sum _{i<j}\tau K_{0}\left(R_{ij,m}/\tilde{\lambda} \right).
\end{equation}

We work only in the limit where $\beta $ is large, which amounts to
taking $L_{z}$ very large and in the mapping onto $2D$ bosons corresponds
to the ground state of the bosons at the absolute zero temperature
($T_{B}=0$). We used $\beta =375$ and discretized the $z$-axis
into $M=75$ planes. 

We used the matrix-squaring method \cite{klemm,ceperley} to
calculate the right action so that we can work with a small number
of planes along the $z$-direction. Working with the primitive action requires
the use of a large number of slicing of the $z$-direction and is very time consuming
\cite{ceperley}. The boundary conditions in the $z$-direction for a system of bosons
are $\mathbf{R}_{i}(\beta )=\mathbf{R}_{j}(0)$, with all permutations
$j=P(i)$ of the indices being allowed. This is what is meant by the
summation over $P$ in Eq. (\ref{Xi}). For small $g$ (a small repulsive
interaction), which corresponds to large $\Lambda$, we expect a
Bose-Einstein condensed phase where permutations are important. This is the
superfluid phase. For small $\Lambda$, repulsion is large and permutations
are rare as the Bose system is in its classical phase which is a Wigner
crystal. These two phases correspond respectively to the FL
liquid and solid phases. The melting line represented by $\Lambda =\Lambda _{m}$
is given approximately by the expression $B_{m}=\mathrm{const.}/T^{2}$. 
Near $T_{c}$ this behavior is more complicated (see below). 

We now discuss the method of simulations of BSCCO : 

Because of the high anisotropy of the BSCCO system one can not use
the simple picture given above. Here, instead, we follow a different
model which can be cast in a form analogous to YBCO. We take the 
Lawrence-Doniach model  \cite{ld} as our starting point for BSCCO. 
It leads to the following
form of the London free-energy for inter-layer (IL) Josephson coupling
\cite{feigelman,ryu}:
\begin{equation}
\Im _{IL}(\mathbf{R}_{i,m})=\frac{d{\phi _{0}}^{2}}{8{\pi }^{3}
{\lambda }^{2}}\left(1+\ln \left(\frac{\lambda }{d}\right)\right)
\left[\frac{{(|\mathbf{R}_{i,m}-\mathbf{R}_{i,m+1}|})^{2}}
{4{r_{g}}^{2}}-1\right],
\label{quadjos}
\end{equation}
 for $|\mathbf{R}_{i,m}-\mathbf{R}_{i,m+1}|<2r_{g}$ , and 
\begin{equation}
\Im _{IL}(\mathbf{R}_{i,m})=\frac{d{\phi _{0}}^{2}}{8{\pi }^{3}
{\lambda }^{2}}\left(1+\ln \left(\frac{\lambda }{d}\right)\right)
\left[\frac{|\mathbf{R}_{i,m}-\mathbf{R}_{i,m+1}|}{r_{g}}-2\right],
\label{linjos}
\end{equation}
 for $|\mathbf{R}_{i,m}-\mathbf{R}_{i,m+1}|>2r_{g}$ , where $d$
is the inter-layer spacing and $r_{g}$ is the healing length defined
by $r_{g}=\gamma d$. For the in-plane (IP) coupling we use:
\begin{equation}
\Im _{IP}(\mathbf{R}_{ij,m})=\frac{d{\phi _{0}}^{2}}{8{\pi }^{2}
{\lambda }^{2}}K_{0}\left(\frac{R_{ij,m}}{\lambda}\right).
\label{logjos}
\end{equation}
In principle one should add to the above interaction an electromagnetic
interaction among the pancake vortices \cite{clem}. The
electromagnetic interaction becomes dominant in the limit of infinite
anisotropy ($\gamma \rightarrow \infty$).
Ryu {\it et al.} \cite{ryu} argue, (see also Ref. \onlinecite{artemenko}) that
for the value $\gamma=50$ for BSCCO the Josephson coupling still
dominates by one order of magnitude over the electromagnetic 
coupling. Their argument goes as follows: Clem \cite{clem} shows that if one has a
straight array of pancake vortices along the $z$-axis, and one pancake
is displaced a distance $R$ in the $x$-direction than the magnetic energy of the
configuration increases by an amount
\begin{equation}
\Delta E(R)=  \frac{d{\phi _{0}}^{2}}{8{\pi }^{2}{\lambda }^{2}}\left(
  {\cal C}+\ln\left(\frac{R}{2 \lambda}\right)+K_0\left(\frac{R}{\lambda}\right)\right).
\end{equation}
where ${\cal C}$ is Euler's constant (=0.5772...). For large $R$ ($R\gg\lambda$), the
modified Bessel function $K_0$ decays exponentially and thus the
energy increases like $\ln(R/\lambda)$. For small $R$ the Bessel
function can be expanded in a power series in $R/\lambda$ 
\begin{equation}
  K_0(R/\lambda)=-\ln(R/2\lambda)(1+R^2/4 \lambda^2+\cdots)-{\cal
    C}+R^2 (1-{\cal C})/4 \lambda^2+\cdots,
\end{equation}
and thus the magnetic energy behaves
like $R^2$ to leading order in $R$ which is the same as the quadratic
behavior of the Josephson energy in Eq. (\ref{quadjos}) above. The ratio
of the coefficients of the quadratic terms in the magnetic and
Josephson energies goes roughly like $\gamma^2(d/\lambda)^2$ (where
$d/\lambda \sim 1/100$ for BSCCO). Thus for anisotropy
$\gamma=50$ we get a factor of 0.25 (a somewhat more precise estimate
\cite{ryu} gives a ratio of about 0.1). Thus the magnetic interaction is
negligible compared to the Josephson interaction for $\gamma=50$. For
samples with $\gamma=200$ these interactions are already comparable. For large
values of $R$ the magnetic interaction increases logarithmically and
the Josephson interaction increases linearly so the magnetic
interaction is always negligible. The key to the estimate given above is
to consider not just two pancake vortices but a whole line with one
displaced pancake. This argument is valid if the deviations of the
vortices from straight lines are not too large. As for the in-plane
interaction Clem showed that a linear array of
pancake vortices gives rise to exactly the same magnetic field at a
distance $R$ away from it as produced by an Abrikosov vortex line. 
Thus Eq. (\ref{logjos}) is consistent with the magnetic interaction of
pancake vortices, again when the FLs do not deviate too much from straight lines.    
 
Equations (\ref{quadjos})-(\ref{logjos}) for BSCCO can be cast in a form
similar to that for the YBCO with the following substitutions:
\begin{equation}
\Lambda =\frac{kT}{\varepsilon _{0}a_{0}}\frac{r_{g}}{d}
\sqrt{\frac{\pi }{2(1+\ln (\frac{\lambda }{d}))}},
\, \, \, \, \, \, \, \, \tau =\frac{2\varepsilon _{0}d}{kT}.
\end{equation}

With these changes the London free-energy functional would look 
like\begin{equation}
\frac{\Im _{IL}(\mathbf{R}_{i,m})}{kT}=\left[
\frac{{(|\mathbf{R}_{i,m}-\mathbf{R}_{i,m+1}|})^{2}-
r_{g}^{2}}{2\Lambda ^{2}\tau }\right],
\end{equation}
 for $(|\mathbf{R}_{i,m}-\mathbf{R}_{i,m+1}|)<2r_{g}$, and 
\begin{equation}
\frac{\Im _{IL}(\mathbf{R}_{i,m})}{kT}=\left[
\frac{{(\mathbf{R}_{i,m}-\mathbf{R}_{i,m+1}})^{2}-
r_{g}^{2}}{2\Lambda ^{2}\tau }\right]-
\frac{(2r_{g}-(|\mathbf{R}_{i,m}-\mathbf{R}_{i,m+1}|))^{2}}
{2\Lambda ^{2}\tau },\label{second}
\end{equation}
 for $(|\mathbf{R}_{i,m}-\mathbf{R}_{i,m+1}|)>2r_{g}$, when now again
all lengths
are measured in units of $a_{0}$. While doing simulations at a fixed
magnetic field $B$ and temperature $T$ , the term $r_{g}^{2}/2\Lambda^{2}\tau$
will remain constant and would drop out of $\Delta E$ term in the
Boltzmann factor. It however needs be considered during the measurement
of the energy. The second term in Eq. (\ref{second}) can be easily
handled at the last stage of the Bisection method (see the next section for a
discussion of the Bisection method). 

We can make use of a reduced temperature variable to make some expressions
look simpler. First, using the fact that the temperature dependence
of $\Lambda $ arises mainly through $\varepsilon _{0}$ and neglecting
the logarithmic corrections, one gets \cite{blatter,margo}
\begin{equation}
\lambda=\frac{\lambda_{0}}{\sqrt{1-T/T_{c}}},\ \ \ \ \varepsilon
_{0}\propto\frac{1}{\lambda^{2}}\propto\left(1-\frac{T}{T_{c}}\right),
\end{equation}

 and hence\begin{equation}
\Lambda \propto \frac{T\sqrt{B}}{1-T/T_{c}}.\end{equation}
 Defining reduced temperature as :\begin{equation}
T_{r}=\frac{T}{1-T/T_c},\end{equation}
 we obtain\begin{equation}
\Lambda \propto T_{r}\sqrt{B}.\end{equation}
 This shows that the equation for the melting line is approximately\[
B_{m}=\mathrm{const.}(1-T/T_{c})^{2}/T^{2}.\]
 Note that some authors \cite{anisotropy} use a temperature dependence
 of $1/\sqrt{1-T^{2}/T_{c}^{2}}$ in $\lambda$, or even  \cite{tinkham}
 $1/\sqrt{1-T^{4}/T_{c}^{4}}$. All these choices
 coincide near $T_c$. The choice of temperature dependence of
 $\lambda$ is not expected to have a significant effect on the results.

\section{Notes on the simulations}

The technique that we use to simulate our system is called Multilevel
Monte Carlo Simulation (MMC) \cite{ceperley}. There are several advantages
in using this technique for the simulation of the FLs over
the usual metropolis Monte Carlo (MC) method. In the discrete model,
we work with $N$ FLs with the $z$-axis discretized into $M$ planes,
thus resulting in $N$ beads in each of the $M$ planes. In the
usual MC method one would displace a few of these beads in a plane
by small random displacements inside a two dimensional box and then
would accept or reject the move based on a probability given by the
Boltzmann factor. 

A big disadvantage of using this technique is that it is difficult
to move beads appreciably from their original positions over a number
of MC steps. The reason for this is that a bead in a plane belonging to
a FL finds itself in a local harmonic potential generated by
the kinetic energy term involving this bead and the beads belonging
to the same FL on either side of the plane. This harmonic potential
becomes stronger and stronger at lower temperatures and magnetic fields.
As a result, in the usual MC simulations beads keep moving around inside
these local harmonic cages and end up sampling only a small part of
the phase space. The other problem with the usual MC method is that there
is no natural easy way of implementing FL cutting. If there
are two FLs twisted around each other and if it is energetically
favorable for them to reconnect each other in such a way as to lower
their free energy then this step should be permitted in the MC
method without regard to the question if this process occurs in reality.
This is so because in the MC simulations phase space is sampled
according to the probability distribution and all one needs is to
generate configurations weighted by the Boltzmann factor, and the path
followed in configuration space has nothing to do with any real dynamics.

These two main drawbacks of the usual MC method are easily overcome
in the MMC technique. First, one moves bigger chunks
of FLs encompassing beads in several planes. This way one can avoid
local harmonic traps. (This is like taking an aerial route to a destination
rather than going through the zigzag maze of roads.) This is much
in the spirit of Fourier space Monte Carlo where one first samples
modes with smaller wave numbers and then move onto higher modes. 

The method of creating new FL configurations is based on the concept of
the conditional probabilities. It is called the Bisection method 
\cite{ceperley} because
one starts sampling beads by iteratively bisecting the Fls. At each stage
of the division, the beads belonging to that stage are moved with some
conditional probability factor $P_i$. It is important to make sure
that the probabilities are chosen in such a way that detailed balance is
satisfied at each stage of the division. One notes that the $P_i$'s may not
be the actual Boltzmann factors for the beads to be moved at different
levels. But what is required is that when all $P_i$'s are multiplied
together, they cancel in such a way so as to leave the correct Boltzmann
factor for the whole move. Thus, inherent in this algorithm is the fact that
a move would finally be accepted only if it has been accepted at each stage of
the Bisection method. The power of this method lies in choosing the
appropriate $P_i$'s. If these conditional probability factors are chosen
judiciously, most of the rejections would take place at the initial stages
of the bisection process when not too much computational effort has been 
spent yet.

The cutting and reconnection of FLs is implemented naturally in a MMC
method: permutation among the 3 or 4 lines chosen to be moved
becomes the first among the many hierarchical steps one goes through
before a move is finally accepted and the position of the beads updated
accordingly. We typically moved a total of 15 to 20 beads distributed over 5
planes. Permutations were sampled by a random walk algorithm through
the space of permutations \cite{nordborg}(see Appendix B). 

In the case of YBCO we worked with a field of $B=4000$ G.
Working in the primitive approximation of the action would require the
use of a smaller value of the dimensionless parameter $\tau $, which
would require slicing of the $z$-direction into a large number of planes. To avoid
this, the matrix-squaring method \cite{klemm,ceperley} has been used to
get the effective action for bigger values of $\tau $. For example
Nordborg and Blatter \cite{nordborg} use a value of $\tau =3.0$
and they work with 100 planes. In this simulation a value of $\tau =5.0$ has been used and
the $z$-axis has been sliced into 75 planes. Choosing a bigger value of $\tau$
by utilizing the matrix-squaring method makes it easier to equilibrate the
system as compared to working with the primitive approximation.

For BSCCO, we did not use the matrix-squaring method because of the 
complications involved due to the few extra terms in free energy which 
contribute depending on whether $R_{ij}$ is smaller or bigger than $r_g$.
Here, we used the natural inter-layer spacing $d$ to calculate the parameter
$\tau$ at different temperatures and then used the MMC technique to efficiently
sample the configurational space.

As mentioned previously, in the present simulations we included only the Josephson coupling. 
This approximation works well with YBCO but it could be less satisfactory for BSCCO because of
 its high anisotropy. For very anisotropic materials
the electromagnetic coupling becomes important \cite{clem,anisotropy}.
As discussed in Section 2, Ryu {\it et al.}
\cite{ryu} estimated that for anisotropy of magnitude $\gamma=50$ the Josephson interaction
still dominates over the electromagnetic interaction by an order of
magnitude, but this will not be the case for $\gamma=200$ which can characterize some samples. Olson
{\it et al.} \cite{olson} discuss how to include the electromagnetic
interaction in a MC simulation, but they only consider the opposite
limit where the Josephson coupling is totally neglected. To our
knowledge there is no satisfactory treatment of both couplings
included on equal footings. We carried out preliminary simulations
which show that the inclusion of the electromagnetic coupling does not shift
the position of the transition line much at a field of $125$ G, thus
supporting our current conclusions. These results will
be reported elsewhere \cite{sandeep}. 

Simulations were usually carried out for $36$ and $64$ FLs. The
decay of the structure factors at the transition temperature becomes
sharper when one uses a larger number of FLs. However,
no appreciable shift in the transition temperature is seen while
working with the smaller ($N=36$) or the larger system ($N=64$).
We did not run our simulations for even larger systems as it becomes
computationally very time consuming. Typical simulation times were
$\sim 3.5\times 10^{6}$ MC steps. Each MC step involved
moving 3-4 lines in 5 planes simultaneously. We usually
averaged over 10-15 realizations of the columnar disorders, though
some results have an average over as many as 20 realizations of
the disorder. 

Columnar disorder is modeled as an array of straight cylindrical wells of typical
radius $r_d$=25-35 \AA placed randomly throughout the cross-sectional region of the
sample and oriented along the $z$-direction \cite{blatter}. Each columnar defect is 
of length $L_z$. The density of the columnar pins can be varied by changing
their number for a given cross sectional area, and the strength is controlled by a positive
dimensionless parameter $\eta $. If a bead happens to wander inside
a columnar well, we include an extra free energy of 
$-\eta \frac{d{\phi _{0}}^{2}}{8{\pi }^{2}{\lambda }^{2}}$.
The defect concentration was taken to be a 20 percent ratio of defects
to FLs which means $B_{\phi }/B=0.2$. The strength of disorder
was set at $\eta \leq 0.5$ for BSCCO. For YBCO, $\eta =0.5$ was found
to be too large in the sense that the transition became too broad
and useful information could not be extracted. Thus, for YBCO we kept $\eta \leq 0.3$. 

Other parameters used for YBCO are $\gamma =5,$ $\lambda _{0}=1500$ \AA,
$\xi _{ab}(0)=15$ \AA, $d=12$ \AA. Parameters for BSCCO are as follows:
$\gamma =50$, $\lambda _{0}=1700$ \AA, $\xi _{ab}(0)=20$ \AA, $d=15$ \AA.

\section{Measured Quantities }

In this section we describe many different physical quantities that were
monitored during the simulation. From the variation of these quantities with
the temperature we can extract important conclusions about the different 
phases of the system.

\subsection{Energy}

In terms of the reduced temperature, the energy

\begin{equation}
E=T^{2}\frac{\partial }{\partial T}\ln (\Xi (\Lambda ,\beta ,N))\end{equation}

can be simply written as \cite{nordborg}:
\begin{equation}
E=T_{r}(S_{1}+S_{2}),\end{equation}
where,
\begin{equation}
S_{1}=\Big\langle\sum _{m,i}\frac{{(|\mathbf{R}_{i,m}-\mathbf{R}_{i,m+1}|})^{2}-
r_{g}^{2}}{2\Lambda ^{2}\tau }\Big\rangle,\, \, \, \, \, \, \, \, \mathrm{for}\,
(|\mathbf{R}_{i,m}-\mathbf{R}_{i,m+1}|)<2r_{g},
\end{equation}
and
\begin{equation}
S_{1}=\Big\langle\sum _{m,i}\left[\frac{{(\mathbf{R}_{i,m}-
\mathbf{R}_{i,m+1}})^{2}-r_{g}^{2}}{2\Lambda ^{2}\tau }\right]-\frac{(2r_{g}-
(|\mathbf{R}_{i,m}-\mathbf{R}_{i,m+1}|))^{2}}{2\Lambda ^{2}\tau }\Big\rangle,
\end{equation}
for  $(|\mathbf{R}_{i,m}-\mathbf{R}_{i,m+1}|)>2r_{g}$,
whereas
\begin{equation}
S_{2}=\Big\langle\sum _{m,i>j}\tau K_{0}\left(\frac{|\mathbf{R}_{i,m}-\mathbf{R}_{j,m}|}
{\lambda }\right)\Big\rangle.
\end{equation}
Any discontinuous jump in $E$ would indicate a FOT. From this discontinuous
jump in energy, $\Delta E$, we can also calculate the jump in entropy 
$\Delta s$,
\begin{equation}
\Delta s = \frac{\Delta E}{T}.
\end{equation}

\subsection{Translational structure factor} 

The translational structure factor $S(\mathbf{Q}_{l_1,l_2})$ is defined as, 

\begin{equation}
S(\mathbf{Q}_{l_{1},l_{2}})=\frac{1}{MN}\Big\langle \sum _{ij,m}e^{\left
(i\mathbf{Q}_{l_{1},
l_{2}}.(\mathbf{R}_{i,m}-\mathbf{R}_{j,m})\right)}\Big\rangle \label{struc}
\end{equation}

where $\langle ...\rangle $ stands for the MC average, 
$\mathbf{Q}_{l_{1},l_{2}}$
is a reciprocal lattice vector and is of the form 
\begin{equation}
\mathbf{Q}_{l_{1},l_{2}}=l_{1}\mathbf{b}_{1}+l_{2}\mathbf{b}_{2}
\end{equation}
and $l_{1}$ and $l_{2}$ are some integers. $\mathbf{b}_{1,2}$
represent the basis vectors of the reciprocal lattice 
\begin{equation}
\mathbf{b}_{1,2}=\frac{2\pi }{a_{0}\sin ^{2}\theta }(\mathbf{e}_{1,2}-
\mathbf{e}_{2,1}\cos \theta) 
\end{equation}
where $\theta =\pi/3$, $a_{0}$ is the nearest neighbor
distance and $\mathbf{e}_{1,2}$ are the unit vectors along the hexagonal
unit cell such that \begin{equation}
\mathbf{e}_{1} \cdot \mathbf{e}_{2}=\cos \theta. 
\end{equation}
 
In the simulations, $l_{1}$ and $l_{2}$ in Eq. (\ref{struc}) are
chosen to be 1 and 0 or 0 and 1. These choices correspond to the first
Bragg peak. We will normally write $S(\mathbf{Q}_{1,0})$ as simply
$S(\mathbf{Q}_1)$. This quantity gives important information about 
the phase of
the system. In an ordered phase where Fls sit on a triangular lattice, 
$S(\mathbf{Q}_1)$ is of the order of $N$. In the disordered phase, it
saturates to almost zero as the system size increases. In the simulations, 
however, there is a problem with measuring $S(\mathbf{Q}_{1})$, especially in 
the presence of disorder. We will return to this point at the end of the
section.

\subsection{Hexatic Structure Factor} 

We use Delaunay triangulation to measure the hexatic order parameter, $\psi_6$,
which is defined as, 
\begin{equation}
\psi _{6}=\langle \sum _{m=1}^{M}\sum _{i=1}^{N}\frac{1}{z_{i,m}}
\sum _{j=1}^{z_{i,m}}e^{(i6\theta _{ij,m})}\rangle ,
\end{equation}
 where $z_{i,m}$ denotes the number of the nearest neighbors of a bead
at position $i$, $m$ and it is 6 for a perfect hexagonal lattice, 
$\theta _{ij,m}$
stands for the bond angle, that is the angle that vector $\mathbf{R}_{ij,m}$
makes with an arbitrary axis. Just like the $S(\mathbf{Q}_1)$, this quantity
too has a large value in an ordered phase and saturates at a finite value
for a system of finite size.

\subsection{Line Entanglement} 

As we allow permutations of FLs, we can define a number $N_{e}/N$
as that fraction of the total number of FLs which belong to loops that are bigger
than the size of a ``simple'' loop. A simple loop is defined as a set of $M$
beads connected end to end, $M$ being the total number of planes.
Loops of size $2M$, $3M$... start proliferating at and above the transition
temperature and in the corresponding 2D boson system  this proliferation is related 
to the onset of the superfluidity. 

\subsection{Line Wandering} 

The transverse FL fluctuations are measured by \begin{equation}
u^{2}(z)=\langle (\mathbf{R}(z_0+z)-\mathbf{R}(z_0))^{2}
\rangle /a_{0}^{2} \label{wander1},
\end{equation}
which is independent of $z_0$. At the transition temperature 
$u^{2}(z)$ undergoes a large increase for large $z$. 

We want to emphasize one important aspect of measuring the translational
structure factor. The usual way of measuring this quantity is by choosing
$\mathbf{Q}_{l_1,l_2}$ corresponding to the first Bragg peak i.e., 
$\mathbf{Q}_1$ in Eq.
(\ref{struc}). Now, it might happen that the configuration of the FLs comes 
close to making an almost perfect hexagonal lattice but its basis vectors are not aligned
along the usual major axes of the rhombically shaped cell encompassing the system.
If we use the reciprocal lattice vector $\mathbf{Q}_{1}$ to measure
the structure factor of such a configuration, we would end up getting
a very small value for $S(\mathbf{Q}_{1})$ and might wrongly conclude
that the system is in a very disordered state. This happened many
times in our simulations; we got a very low value of the translational
structure factor while the hexatic order was indicating a high degree
of orderliness in the FL lattice. To remedy this situation
the translational structure factor is measured at 60 different angles,
1 degree apart, and choose that number which gives the largest possible
value of the $S(\mathbf{Q}_{1}(\alpha ))$ corresponding to some
angle $\alpha $. After implementing this technique we find that
$S(\mathbf{Q}_{1})$ and $\psi_6$, which differ so much initially,
almost follow each other.

\section{Results}

1. For YBCO, simulations were carried out at a magnetic field of $B=4000$ G.
At this field we have $2a_{0}\approx \lambda_0 $. By the argument given
in Ref. \onlinecite{nordborg}, we conclude that our results should hold
qualitatively for any $B$ such that $a_{0}<\lambda $. We checked
our simulation results against those by Nordborg and Blatter \cite{nordborg}.
First we verified that our code was working fine by comparing our results
against those given in Ref. \onlinecite{nordborg}. Working in the limit
of $L_{z}\rightarrow \infty $, we kept $\tau $ fixed as
$\Lambda$ was increased. We got a sharp transition at $\Lambda =0.0605$. A
jump was also seen in the energy as defined in the previous section and was
found to be $0.013k$, again the same as in Ref. \onlinecite{nordborg}. 

\begin{figure}[ptb]
\begin{center}
\includegraphics{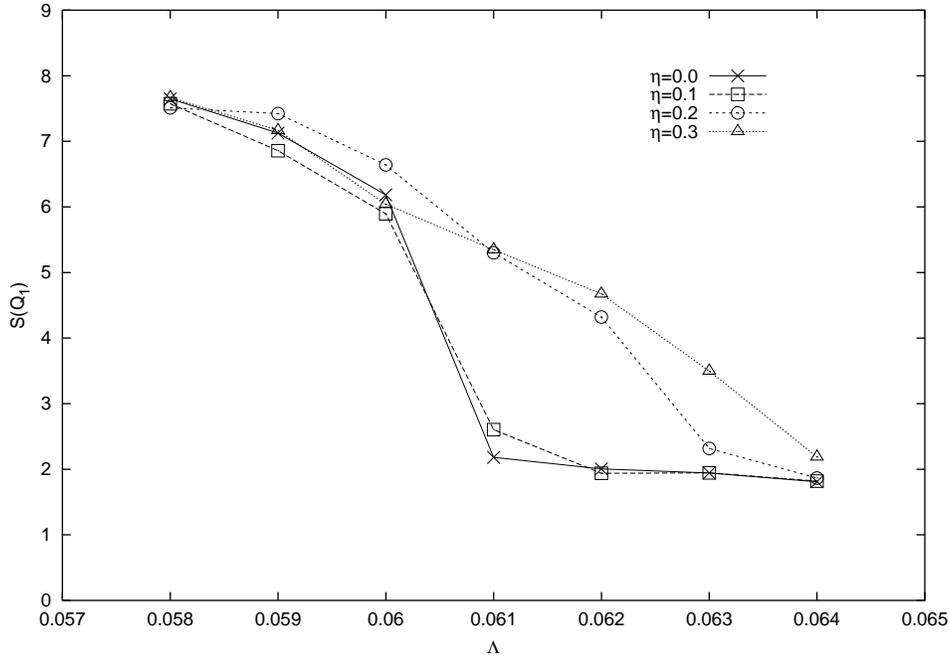}
\end{center}
\caption{YBCO: Structure Factor at the first Bragg peak as a function of
 $\Lambda$ at 
various disorder strengths ($\eta$'s) for $N=36$. A clear shift in transition 
point towards higher $\Lambda$'s is seen for $\eta\geq0.2$}%
\label{all1}
\end{figure}

\begin{figure}[ptb]
\begin{center}
\includegraphics{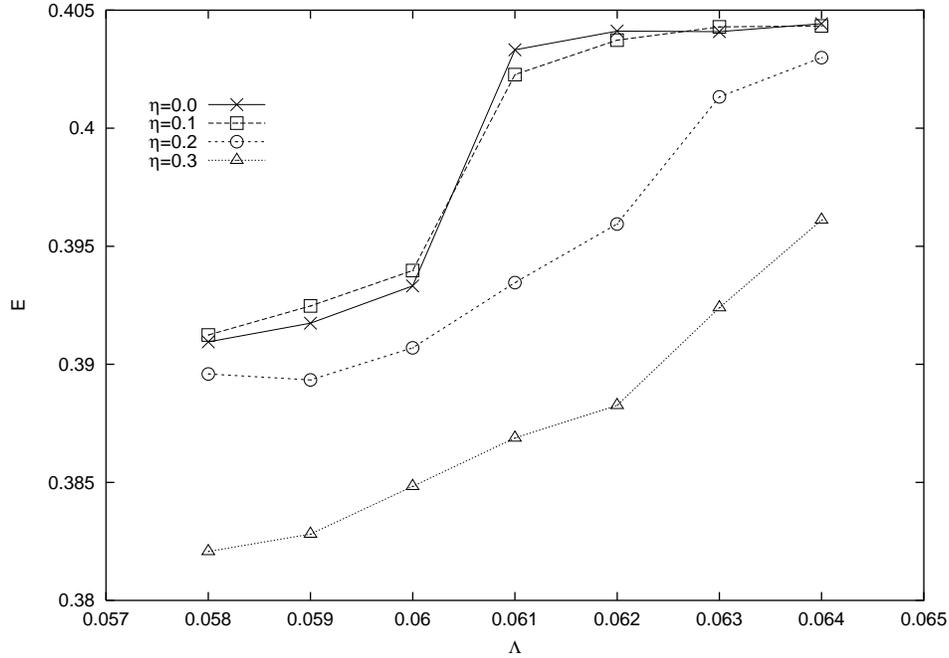}
\end{center}
\caption{YBCO: Energy as a function of $\Lambda$ for $N=36$. 
A jump in the energy is
seen up to $\eta=0.1$. For $\eta\geq 0.2$, the rise in the energy is soothed 
out.}
\label{all2}
\end{figure}

The effect of introducing columnar disorder on the structure
factor at the first Bragg peak 
is shown in Fig. \ref{all1}. In Fig. \ref{all2} we depict the jump in the energy for
a pure as well as for a disordered system. 

We note that at lower strengths of the disorder, the melting line is almost
unaffected and the transition takes place at $\Lambda \approx 0.0605$.
However, as we increase the strength of the disorder,
the melting transition shifts towards higher values of $\Lambda$
which means that the melting line shifts towards higher temperatures
and/or higher magnetic fields. For $\eta $ up to $0.1$ no change
is seen in the melting curve or in the jump in the free energy functional.
However, at $\eta $ equal to $0.2$ the structure factor comes down
at around $0.0625$ and, the jump in the energy becomes gradual. This
finding is in agreement with several experimental studies where a change
in the irreversibility line is seen with the introduction of columnar
disorder \cite{konczykowski,civale,samoilov,paulius,olsson}. 

\begin{figure}[ptb]
\begin{center}
\includegraphics{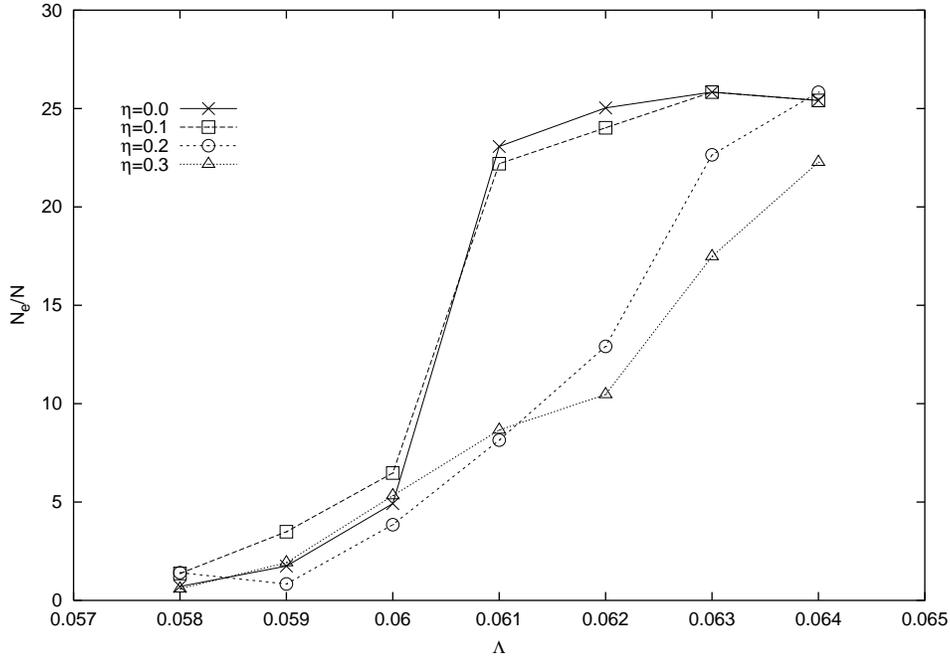}
\end{center}
\caption{YBCO: Fraction of entangled FLs as a function of  $\Lambda$. 
For $\eta\leq 0.1$ a sharp jump in the FL entanglement 
is seen. For $\eta\geq 0.2$, the rise of $N_e/N$ with
$\Lambda$ becomes gradual as compared with the clean system.}
\label{all3}
\end{figure}

Next, we look at the FL entanglement. $N_{e}$ denotes the number of 
FLs which do not form a simple loop. By looking at the $N_{e}/N$ vs. $\Lambda $
graph in Fig. \ref{all3}.
it is clear that the entanglement is suppressed
by as much as almost one order of magnitude just after and in the
vicinity of the original transition line at $\Lambda =0.0605$. This
result is expected, as it is well known that point disorder helps in
line wandering and entanglement while the columnar disorder has just
the opposite effect \cite{goldschmidt} since a nearby FL is induced to
align along the columnar defect, and thus its transverse fluctuations are reduced. 
Unfortunately, there have been few studies on columnar disorder in YBCO. 
Sen {\it et al.} \cite{sen}
work with vanishingly small magnetic field and infinite pinning strengths,
in the vicinity of the lower branch of the melting line. 

\begin{figure}[ptb]
\begin{center}
\includegraphics{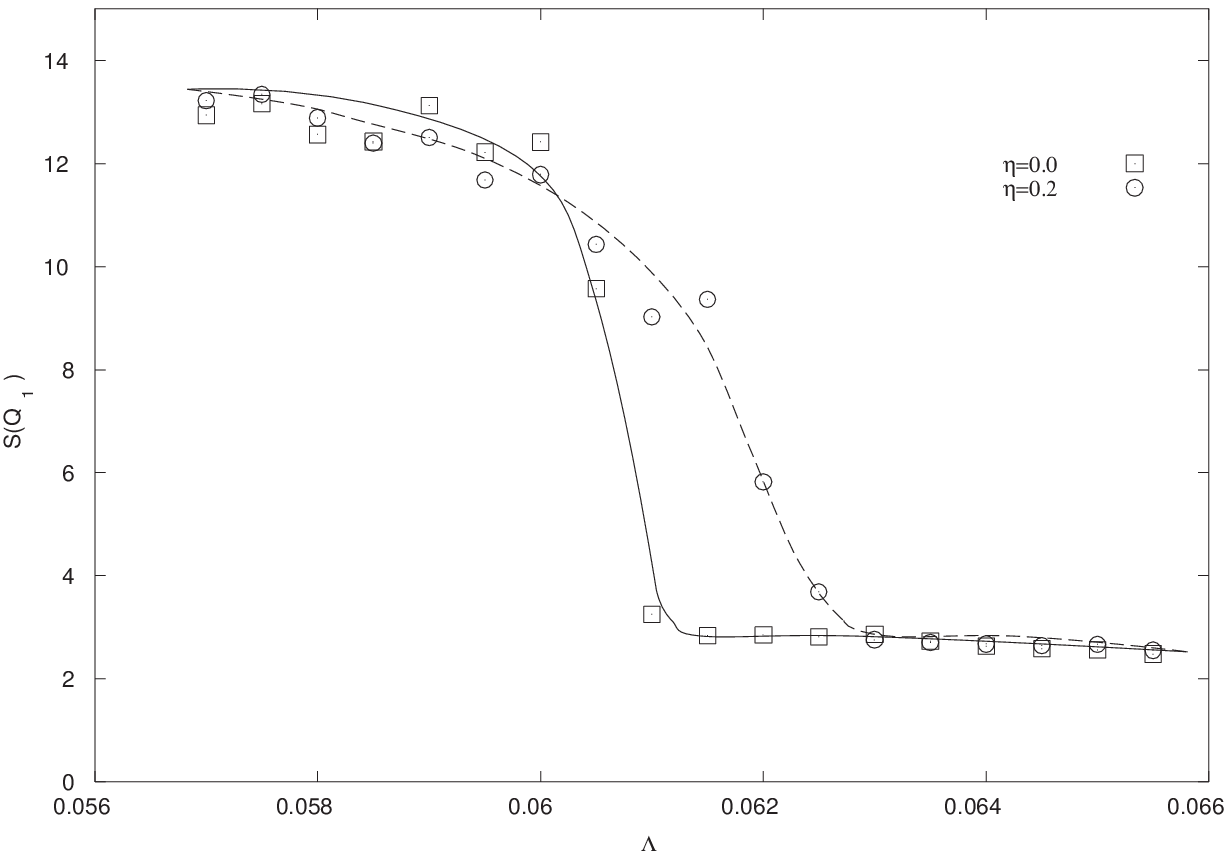}
\end{center}
\caption{YBCO: Structure Factor at the first Bragg peak vs. $\Lambda$ for 
$N=64$. Smooth lines are provided as a visual aid. Transition for $\eta=0.0$ and $\eta=0.2$ are seen almost at the same
$\Lambda$'s as with $N=36$.}
\label{64ybco}
\end{figure}

Figure \ref{64ybco} shows the effect of columnar disorder on a system
of $64$ FLs in YBCO.
This graph shows a transition at almost the same $\Lambda $ as in Fig.
\ref{all1}. This is a confirmation that finite size effects are
not important in our simulations.

\begin{figure}[ptb]
\begin{center}
\includegraphics{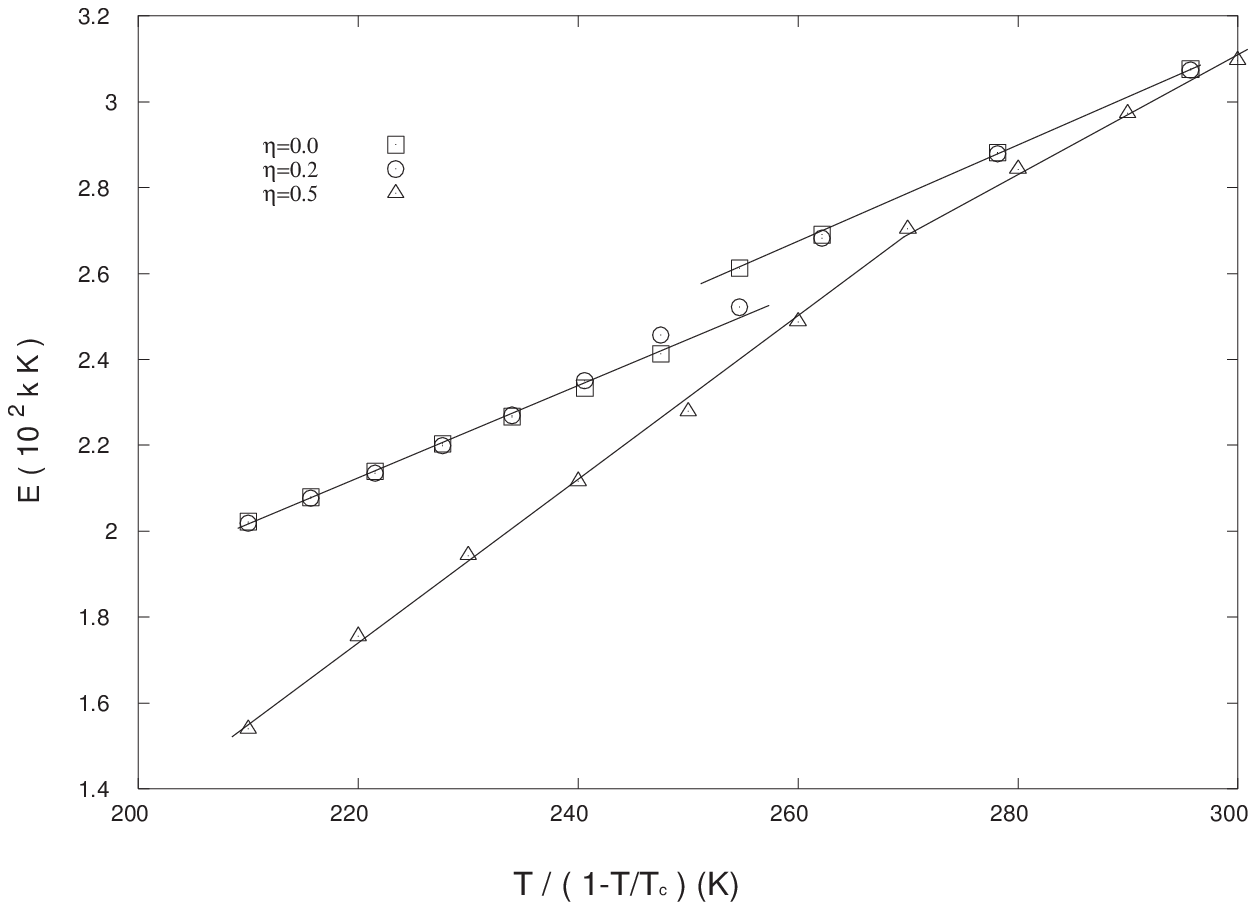}
\end{center}
\caption{BSCCO: Energy as a function of the reduced temperature for 
$\eta=0,0.2,0.5$ at $B =125$ G for $N=36$. Lines were added to help
visualize the energy jumps.
A discontinuous change in energy is seen at $T_r=250,260$ K ($T\approx
66.2, 66.9$ K) for $\eta=0,0.2$ respectively. For $\eta=0.5$ only a
change in slope can be seen at $T_r \approx 268$K ($T \approx 67.4$ K)
instead of a discontinuous jump.}
\label{36energy}
\end{figure}

\begin{figure}[ptb]
\begin{center}
\includegraphics{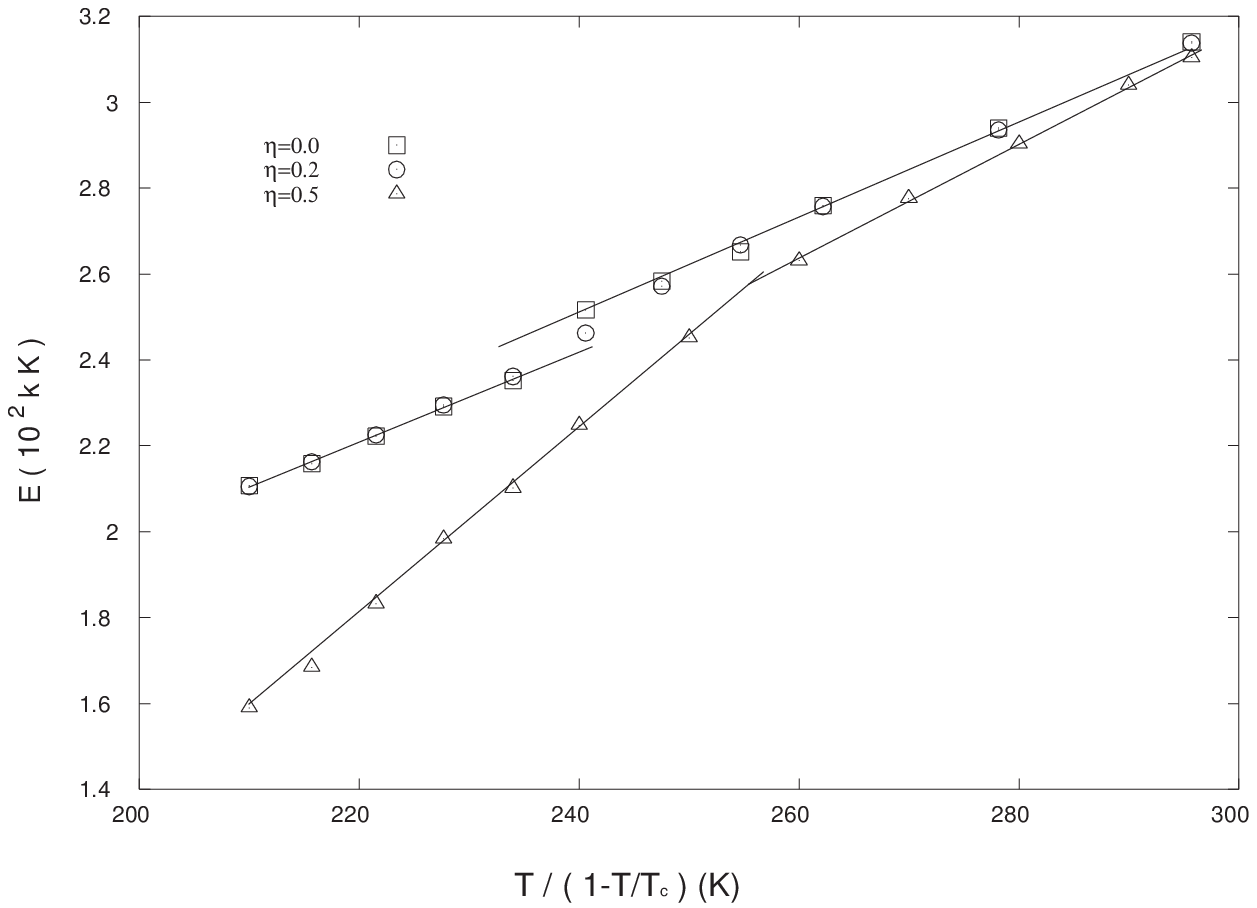}
\end{center}
\caption{BSCCO: Energy as a function of the reduced temperature for 
$\eta=0,0.2,0.5$ at $B=125$ G for $N=64$. A jump in energy can be seen 
at $T_r\approx 238, 242$ K ($T\approx 65.3, 65.6$ K) for $\eta=0,0.2$
respectively. Again no discontinuity is observed for $\eta=0.5$, but
only a change in slope at $T_r \approx 255$ K ($T \approx 66.5$ K).}
\label{64energy}
\end{figure}

2. For the BSCCO system the $E$ vs. $T$ graphs for different $\eta$
values are depicted in Fig. \ref{36energy} (36 FLs) and
Fig. \ref{64energy} (64 FLs).

For the pure system ($\eta=0$) we see a sharp jump in the energy at exactly the
point where the translational as well as hexatic structure factors
sharply decline. This jump in energy is a clear signature of a first
order transition. From the energy jump we can compute the jump in
entropy ($\Delta s$).
The jump in energy at the transition is around $\Delta E=12.0k$ K
per vortex per layer, which gives $\Delta s=\Delta E/T_{m}=0.19k$
which is small compared to the experimental value of $\Delta s=0.40k$
at $B\approx 125$ G. However, these values for $\Delta s$, $T_{m}$ and $B$
are in qualitative agreement with the same system studied using a different
model \cite{hu}. 

\begin{figure}[ptb]
\begin{center}
\includegraphics{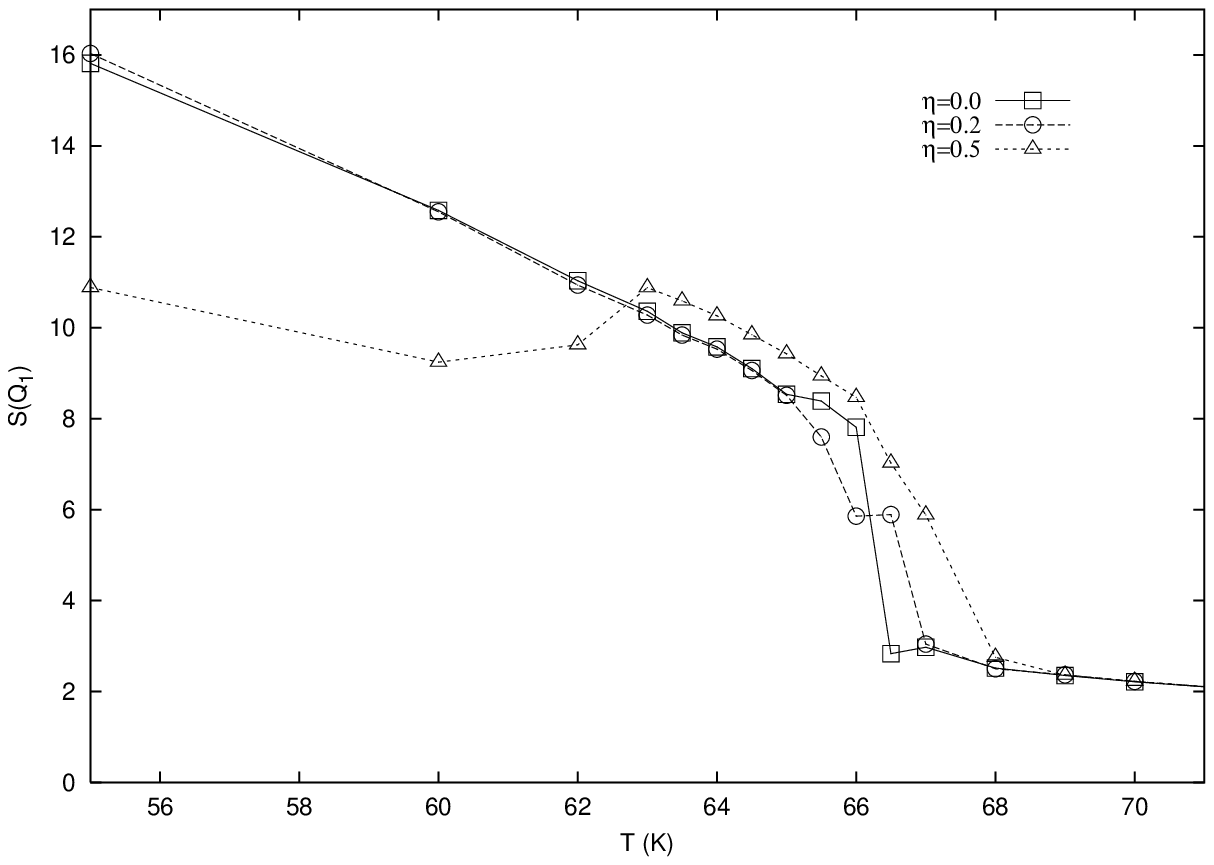}
\end{center}
\caption{BSCCO: Structure factor at the first Bragg peak  vs. temperature  
at $B=125$ G for $N=36$. No shift 
in transition point is seen for $\eta=0.2$. In the presence
of columnar pins with $\eta=0.5$, the transition temperature increases from $66$ K
to $68$ K. Also, we can see that $S(\mathbf{Q}_1)$ starts to rise close to
$60$ K }
\label{36stru}
\end{figure}

\begin{figure}[ptb]
\begin{center}
\includegraphics{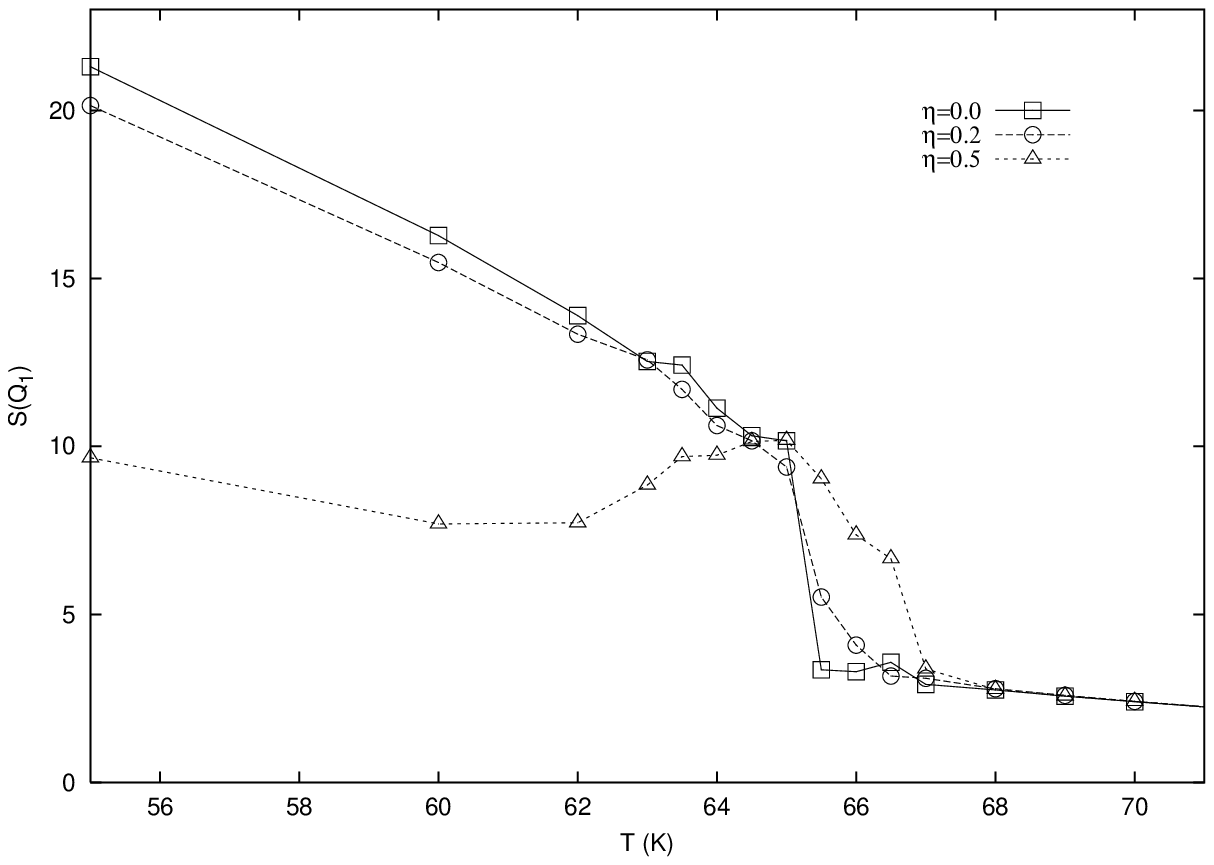}
\end{center}
\caption{BSCCO: Structure factor at the first Bragg peak vs. temperature at 
$B=125$ G for $N=64$. For $\eta=0.2$ no significant change in the
structure factor is observed. In the presence
of columnar pins with $\eta=0.5$, transition temperature increasing from $65$ K
to $67$ K. Also, we can see that $S(\mathbf{Q}_1)$ starts to rise close to
$61$ K }
\label{64stru}
\end{figure}

\begin{figure}[ptb]
\begin{center}
\includegraphics{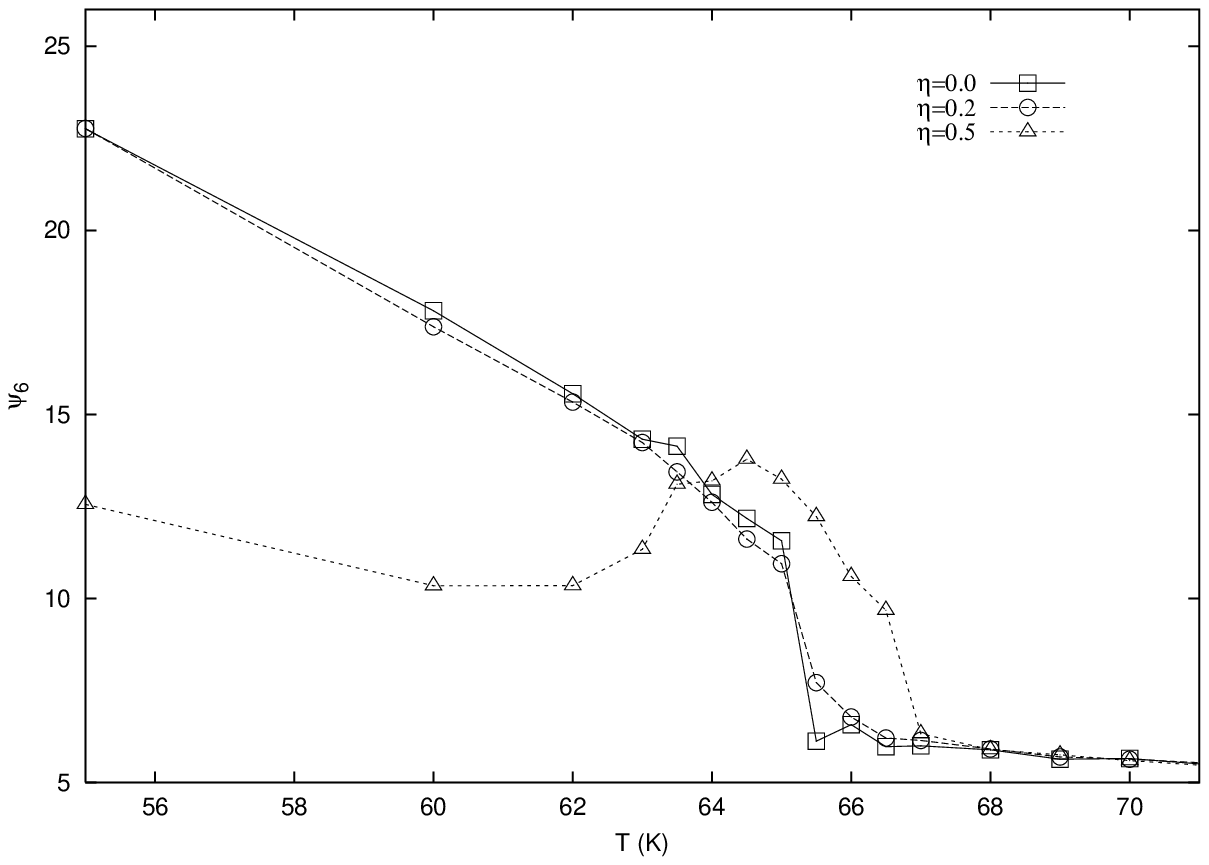}
\end{center}
\caption{BSCCO: Hexatic order vs. temperature at $B=125$ G for $N=64$. When compared
with the previous figure, it can be seen that $\psi_6$ almost follows
 $S(\mathbf{Q}_1)$.}
\label{64hexa}
\end{figure}

Introduction of the columnar defects of a finite strength shows some
interesting effects. We put columnar pins at random positions with a 
concentration fixed at 20 percent of the FLs. We study the BSCCO system
for pins of low strength ($\eta =0.2$) as well as for a higher strength
($\eta =0.5$). Columnar disorder of strength up to $\eta =0.2$
appear to have little effect on the system . This can be seen
from the energy vs. temperature graph in Figs. \ref{36energy} and
\ref{64energy}, where the curve for the pure system and
the one in the presence of low disorder strength ($\eta=0.2$) fall on top of
each other, except in the transition region where the jump in case of the
system with columnar pins is more gradual. Also, from
Figs. \ref{36stru} and \ref{64stru} we see that the $S(\mathbf{Q}_1)$
is not much affected in the presence of the columnar disorder of strength
$\eta=0.2$.  The same is true for the hexatic order depicted in
Fig. \ref{64hexa}. This
means that for the lower strengths of the disorder, the 
vortex-vortex interaction is dominant in the range of the temperature where
we carried out simulations. As a result the first order vortex-lattice 
melting transition (FOT) line is almost not affected.

The nature of FL melting changes, however, when the strength of the 
disorder is increased to $\eta=0.5$ as seen in Figs. \ref{36stru} and \ref{64stru}. 
We see that at the lower temperatures the order parameter is suppressed. As
we increase the temperature, the order parameter starts to rise and
joins the melting curve of the pure system and then falls along it
at even higher temperatures. Fig. \ref{64hexa} indicates that the hexagonal 
structure factor also shows a similar behavior.
This clearly shows that the FLs
start to disengage from the pinning centers as they wiggle more. 
At higher temperatures, columnar pins with the strength
chosen here are not effective in pinning the vortex system. This becomes 
clear from the fact that the
melting curve for the pure system and the one with columnar disorder
join each other at a temperature below the FOT. Further interpretation
and a possible explanation of this phenomenon will be discussed in
more detail in the Conclusions section below. This convergence of the
curves for the pure and disordered cases is also borne out in the
$E$ vs. $T$ graphs in Fig. \ref{36energy} and Fig. \ref{64energy}. 
Initially at low temperatures there is a big difference
in energies of the systems with no disorder ($\eta =0.0$) and high
disorder ($\eta =0.5$). However as the temperature is increased,
the two curves come closer and finally merge together in the liquid
phase. The jump in energy  in presence of columnar defects can still be seen 
in the $E$ vs. $T$ graph for
a disorder strength of $\eta =0.2$. This tends to suggest that
the FOT in a BSCCO system is not affected by disorders kept at as many
as 20 percent of the total number of lines used in the simulations
as long as we keep the strength of the disorder less or equal to $\eta =0.2$.
For $\eta =0.5$, the rise of energy with temperature becomes much
more gradual and we do not see any discontinuous jump in energy at
any temperature. On the other hand an abrupt change in slope of the
energy vs temperature graph is observed, suggesting a discontinuity in
the specific heat characterizing a second order transition.

\begin{figure}[ptb]
\begin{center}
\includegraphics{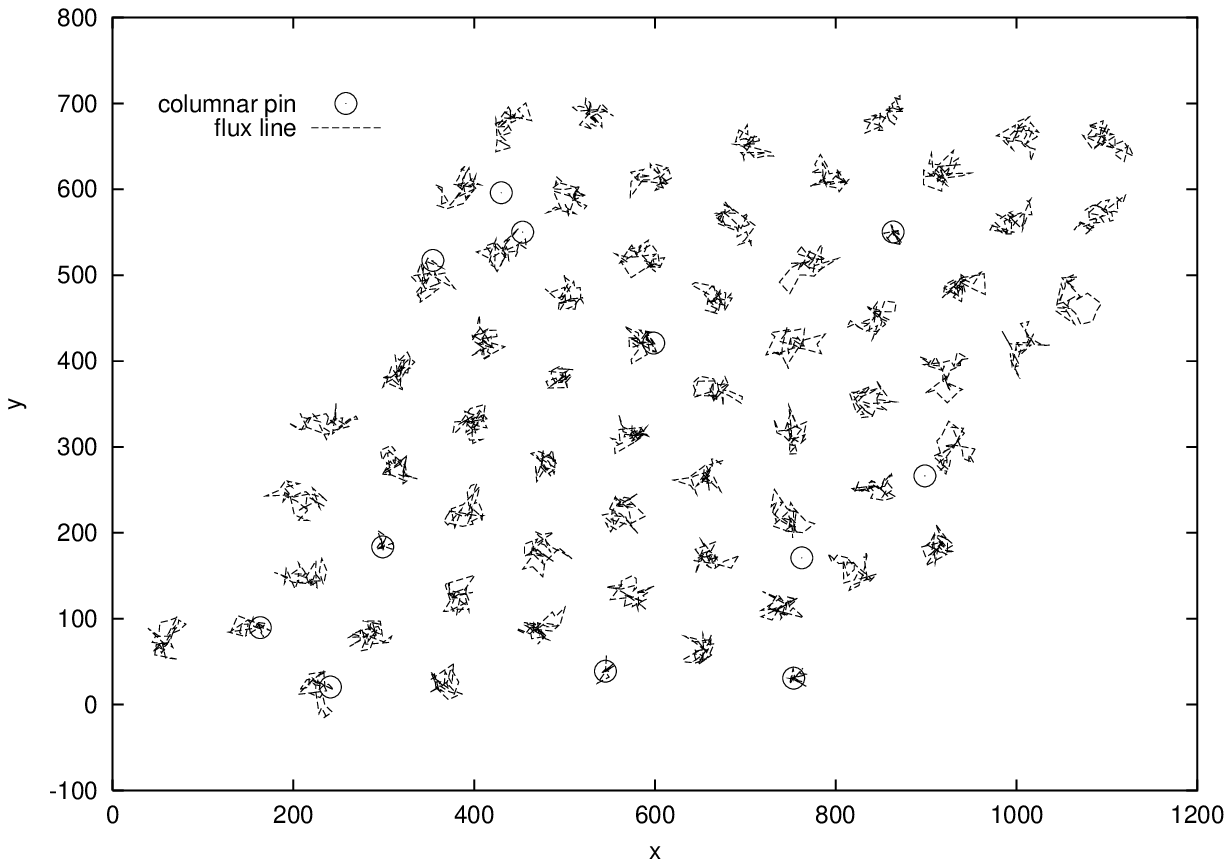}
\end{center}
\caption{BSCCO: A typical configuration in the solid phase (low temperature) 
for $N=64$ FLs
for $B_{\phi}/{B}=0.2$. FLs have been projected onto one plane.}
\label{64low}
\end{figure}

\begin{figure}[ptb]
\begin{center}
\includegraphics{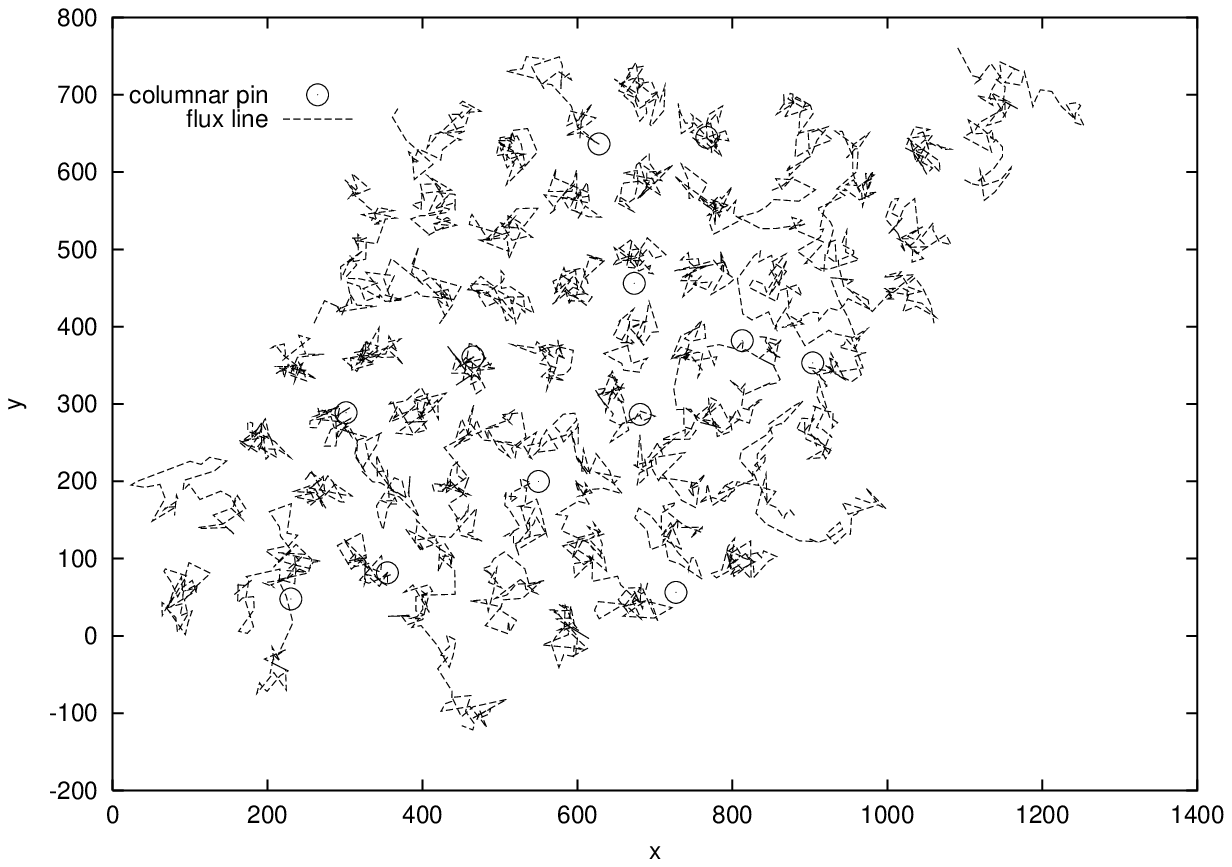}
\end{center}
\caption{BSCCO: A typical configuration of FLs in the liquid phase
(high temperature) for $N=64$ and  $B_{\phi}/{B}=0.2$. FLs have been 
projected onto a single plane. Columnar defects are not drawn to scale.
Some FLs on the boundary do not seem to make loops. That is only 
because virtual images of FLs outside the cell are not shown.}
\label{64high}
\end{figure}

In Fig. \ref{64low} and Fig. \ref{64high}, the snapshots of FLs, projected on a
plane, at temperatures less than the transition temperature and at
a temperature bigger than the transition temperature are shown. 
From Fig. \ref{64low} it is easily seen that at low temperatures most columnar
defects have captured FLs. Also, the FLs make simple loops and
have cleverly set themselves so as to make a hexagonal lattice and
yet occupy as many defects as possible. The transverse fluctuations
of the trapped FLs are greatly suppressed at low temperatures.
At higher temperature beyond the transition point, we see that columnar
defects are not occupied any more and a lot of FLs are entangled.

Inspection of snapshots like Fig. \ref{64low} gives support to the assertion
of Sen {\it et al.} that the Bose glass consists of patches of ordered
regions with only short range positional and orientational order. This
phase is different from the Bragg-glass in systems with point pins
which is characterized by quasi-long-range order.

\begin{figure}[ptb]
\begin{center}
\includegraphics{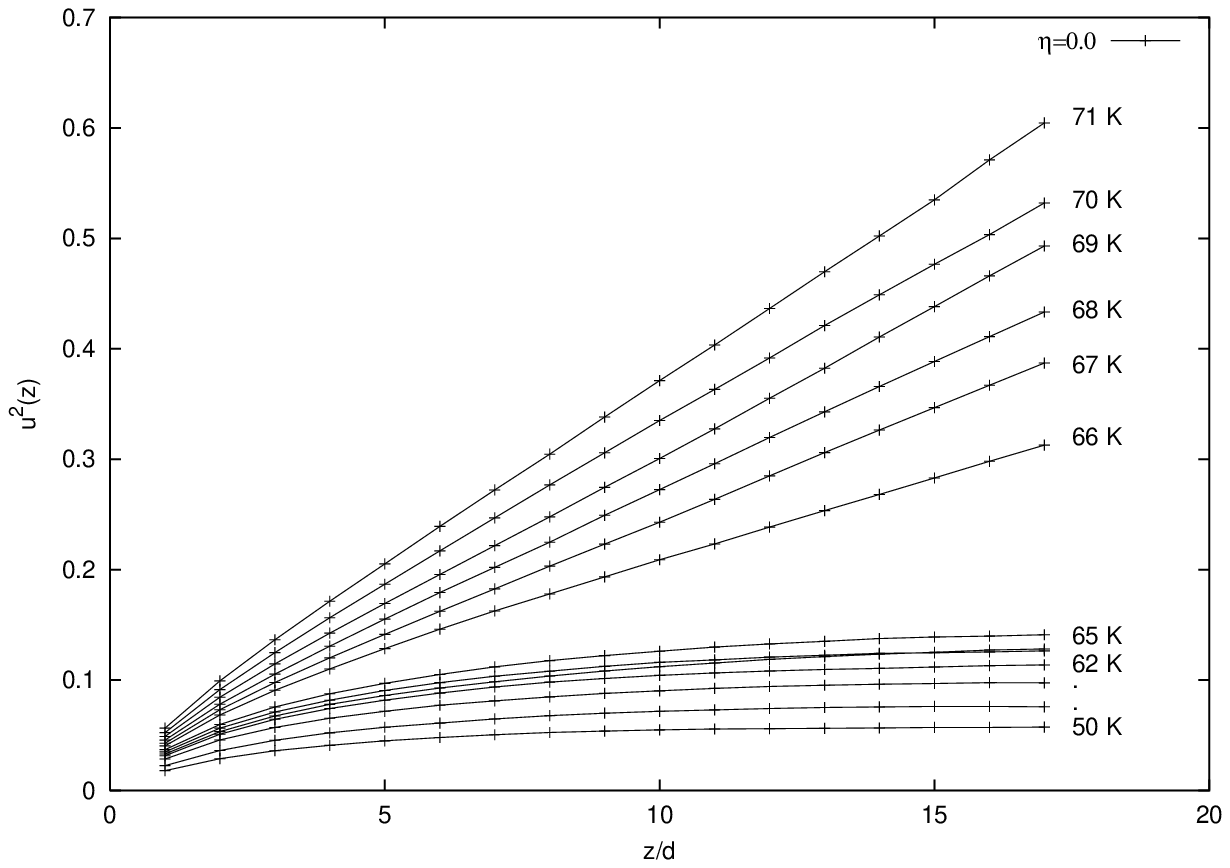}
\end{center}
\caption{BSCCO: Line wandering along the $z$-direction at $B=125$ G 
for $N=64$. Here $d$ is the distance between adjacent planes.
A large increase in line wandering  occurs at $T\approx 65$ K.}
\label{64wander}
\end{figure}

Figure \ref{64wander} shows the mean squared displacements of
the FLs in the $z$-direction at different temperatures for a
pure system.
At lower temperatures $u^2(z)$ saturates for large $z$
but in the liquid state it grows linearly \cite{koshelev}. A large gap in $u^2(z)$
for large values of $z$ seen at the transition temperature signals 
the onset of the entanglement
of the FLs.  

\begin{figure}[ptb]
\begin{center}
\includegraphics{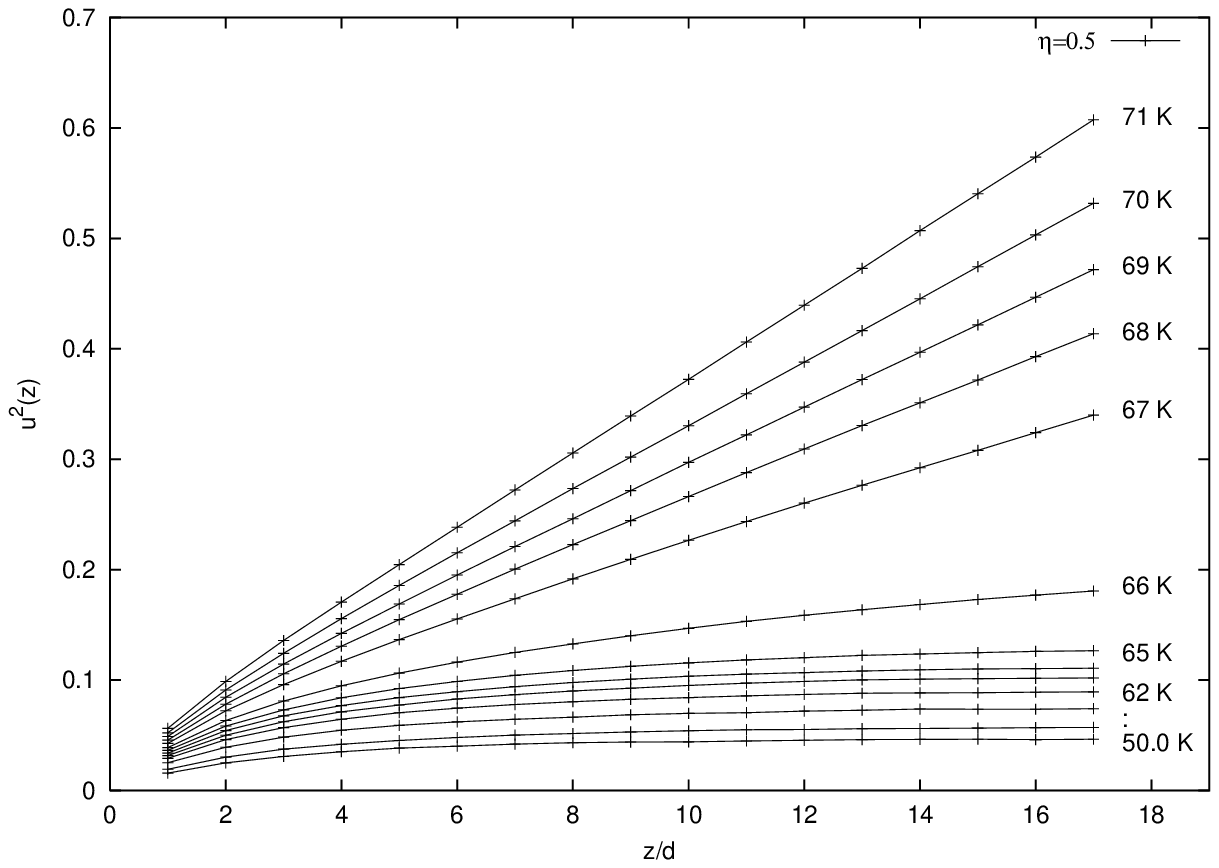}
\end{center}
\caption{BSCCO: Line wandering along the $z$-direction at $B=125$ G for 
$N=64$ in presence of
disorder of strength $\eta=0.5$. The big jump in $u^{2}(z)$ has moved
to $T=66$ K now.}
\label{64wander_col}
\end{figure}

In the presence of the columnar disorder of the strength $\eta =0.5$
we see from Fig. \ref{64wander_col} that $u^2(z)$ at temperatures
less than the transition temperature is suppressed compared to the
corresponding $u^2(z)$ in the pure system. 
This result is in agreement
with the findings in Ref. \onlinecite{goldschmidt}. Also the big gap 
in $u^2(z)$ occurring at the melting transition has moved towards a higher temperature
signaling a shift in the position of the melting transition in the presence 
of columnar defects.

\section{Conclusions}

For lower strengths of disorder ($\eta \leq 0.1$ for YBCO and $\eta \leq 0.2$
for BSCCO) no appreciable shift in the melting line was seen. For these 
strengths, a sharp drop in the translational and hexatic structure factors 
takes place at the transition. Also, the jump in energy at the transition 
is not affected much as compared with the case when there is 
no disorder present. This suggests that for smaller concentrations of the
columnar defects the transition still remains first order.
This result is in agreement with Ref. \onlinecite{khaykovich} where even 
with the introduction of the columnar
disorder no shift in transition temperatures is seen as long as the 
concentration of columnar disorder introduced is small.

It is found that for YBCO as well as BSCCO, the melting transition shifts 
towards higher values of temperature and magnetic field when random
disorder is introduced provided its strength exceeds a threshold which
is different for YBCO and BSCCO. The size of the shift, for a given 
concentration of defects, depends on the strength of the disorder. For YBCO, 
a considerable shift in the
melting line towards higher temperatures and magnetic fields was seen 
for $\eta =0.2$ and $\eta =0.3$. This shift was bigger for
$\eta =0.3$ than for $\eta =0.2$. Similarly, for BSCCO, a large shift in 
the melting line towards a higher temperature was found for $\eta =0.5$.
These results are in tune with numerous experimental findings \cite{
civale,konczykowski,paulius,khaykovich} as well 
as theoretical prediction \cite{goldschmidt} where the 
irreversibility line is seen to shift towards higher temperature and magnetic
fields in the presence of columnar defects. The qualitative reason for this 
effect is that due to the interaction with the columnar defects the transverse 
thermal fluctuations of the FLs are reduced and thus the melting 
transition, as determined from the Lindemann criterion, takes place at
at a higher temperature.

At $\eta =0.5$ the jump in energy is not discernible any more. 
Instead, a change of slope corresponding to a specific heat
discontinuity is observed. This
means that the transition is probably not of first order in the
presence of columnar disorder of higher strengths, but it is rather a
continuous (second order) transition.

The most dramatic outcome of this study for BSCCO is that for some
values of the applied field and defect strength both the 
translational and hexatic
structure factors start to rise at a certain temperature as the
transition is approached from the lower temperature side. This is an 
unusual result, that to our knowledge, has not been seen in previous 
simulations.
This fact can be explained as follows. At low temperatures, the free energy 
is dominated by energy effects rather than entropy considerations, and pinning
effects are dominant. As a result most columnar defects capture 
a FL, while the rest of the FLs adjust themselves in positions such as to minimize
the free energy.  However, as the temperature is increased, FLs 
start to decouple from the defect sites, which allow the vortices pinned at 
interstitial positions to move themselves into a more ordered arrangement, 
thus increasing the structure factors compared to the situation at lower
temperatures. As was mentioned before, due to its high anisotropy, FLs in BSCCO 
behave more like a collection of two dimensional pancakes rather than rigid rods. 
These pancakes are comparatively more difficult to get pinned all at once by
a columnar defect. On the other hand, FLs are much stiffer in YBCO and it
is easier for a columnar defect to capture a FL all along its length. It can
be shown (Eq. (9.49) in Ref. \onlinecite{blatter}) that the depinning temperature 
for FLs, $T_{dp}$, is inversely proportional to the anisotropy parameter 
$\gamma$. Thus, we expect FL depinning to occur at comparatively smaller
temperatures in BSCCO than in YBCO. Since the melting temperatures are
not that different between these materials at the fields we consider,
the depinning process for BSCCO occurs further below the melting temperature than in YBCO.
This allows the structure factor to increase above the depinning temperature before its ultimate 
decline at the melting transition.
For YBCO, the depinning takes place close to the melting temperature, 
and it is difficult to detect any rise in the structure factors, 
especially for small systems, because it is masked by the decrease in order due to 
the stronger thermal fluctuations.

In principle, The rise of the order parameters in the presence of low amount
of columnar defects as the melting transition temperature is approached could be
observed experimentally, if the appropriate parameters are tuned correctly. In small
angle neutron scattering (SANS) \cite{cubitt,lee} one can measure the integrated intensity
over a Bragg peak of wave vector ${\bf Q}_1$ which is proportional to the translational 
structure factor $S({\bf Q}_1)$ measured in the simulation. Another commonly used technique is muon 
spin rotation \cite{lee} which gives information about the width of the magnetic 
field distribution in the sample. This is not directly proportional to the structure 
factor at a given ${\bf Q}_1$, but is rather given by
\begin{equation}
\langle\Delta B^2\rangle(T)=B^2 \sum _{\mathbf{Q}\neq 0}\frac{\exp
  (-\mathbf{Q}^2 \langle u^2\rangle/2)}
{[1+\lambda ^{2}(T)\mathbf{Q}^{2}]^2}, 
\end{equation}
where $\langle u^2\rangle$ is the mean square deviation of vortices
from their average position. It is possible to measure this quantity
in the simulations but this was not done it in the present work. This
quantity might not show the unusual rise describe above since it is
dominated by $\langle u^2 \rangle$ which is very likely monotonically
increasing with temperature yilding a monotonically decreasing line width.
Thus in order to look for the effect observed in this paper we suggest
using the SANS technique to look at a BSCCO sample with columnar pins described by a matching
field of about 25 G and an applied field of about 125 G. It is not clear
to us what is the corresponding $\eta$ parameter describing the pinning
strength of the experimental defects. According to Blatter {\it et
al.} \cite{blatter}  $\eta$ lies in the range 0.1-1.  

\section{Acknowledgements}
This work is supported by the US Department of Energy (DOE), grant No. DE-FG02-98ER45686.

\appendix
\section{Energy sum over the images}

Consider a rhombically shaped region with side $L$ and angle $\theta ,$
unit vectors are $\mathbf{e}_{1}$, $\mathbf{e}_{2}$, with 
$\mathbf{e}_{1}\cdot\mathbf{e}_{2}=\cos \theta$. In practice we took 
$\theta =60^{\circ }$
but we leave the discussion here more general. The Green's function
$G_{0}$ which describe the 2D coulomb interaction between one vortex
and another including all its images, as is implied by the periodic
boundary conditions is given by the solution to London's equation (see e.g. Ref.
\onlinecite{tinkham})
\begin{equation}
(1-\lambda ^{2}\nabla ^{2})G_{0}(\mathbf{R},\lambda )=\lambda ^{2}\delta (\mathbf{R}),
\end{equation}
 with the parameter $\lambda $ setting the scale for the range of
the interaction. The solution is given by
\begin{equation}
G_{0}(\mathbf{R},\lambda )=\frac{2\pi \lambda ^{2}}{L^{2}\sin \theta }
\sum _{\mathbf{Q}}\frac{\exp (i\mathbf{Q}\cdot \mathbf{R})}
{1+\lambda ^{2}\mathbf{Q}^{2}},\label{expansion}
\end{equation}
with
\begin{equation}
\mathbf{R}=R_{1}\mathbf{e}_{1}+R_{2}\mathbf{e}_{2},
\, \, \, \, \, \, \mathbf{Q}=n_{1}\mathbf{b}_{1}+n_{2}\mathbf{b}_{2,}
\end{equation}
 where $\mathbf{Q}$ runs over all reciprocal lattice vectors spanned
by\begin{equation}
\mathbf{b}_{i}=\frac{2\pi }{L\sin ^{2}\theta }(\mathbf{e}_{i}-\mathbf{e}_{j}\cos \theta ),
\end{equation}
 for $(i,j)=(1,2),(2,1)$ . Substituting in Eq. (\ref{expansion}) we obtain
\begin{equation}
G_{0}(\mathbf{R},\lambda )=\frac{\sin \theta }{2\pi }\sum _{n_{1}=
-\infty }^{\infty }\sum _{n_{2}=-\infty }^{\infty }
\frac{\exp [i\frac{2\pi }{L}(n_{1}R_{1}+n_{2}R_{2})]}{L^{2}\sin ^{2}
\theta /(2\pi \lambda )^{2}+n_{1}^{2}-2n_{1}n_{2}\cos \theta +n_{2}^{2}}.
\end{equation}
 We are now going to carry out the summation over $n_{1}$ analytically.
This can be done by using the formula
\begin{equation}
\sum _{n=-\infty }^{\infty }f(n)=-\sum (\mathrm{residues}\, 
\, \mathrm{of}\, \, \pi f(z)(\cot \pi z-i)\, \, \mathrm{at}\, 
\, \mathrm{poles}\, \, \mathrm{of}\, \, f(z)).
\end{equation}
 We have subtracted the constant $i$ from $\cot \pi z$ to ensure
that $|zf(z)(\cot \pi z-i)|\rightarrow 0$ on the contour of integration
when $z$ has a large negative imaginary component. The contour of
integration is a square with sides parallel to the real and imaginary
axes with the origin in the middle, in the limit that its size goes
to infinity. In our case
\begin{equation}
f(z)=\frac{e^{ixz}}{(z-\beta -i\gamma )(z-\beta +i\gamma )},
\end{equation}
 with appropriate values of $\beta $ , $\gamma $ , and $x$ . This
function has two simple poles, and the residues can be easily evaluated.
The final answer becomes (relabeling $n_{2}$ as $n$ ) 
\begin{equation}
G_{0}(\mathbf{R},\lambda )=\frac{\sin {\theta }}{2}\sum _{n=
-\infty }^{+\infty }\frac{\cos (t_{2}n-2\pi \beta _{n})\sinh 
(\gamma _{n}t_{1})+\cos (t_{2}n)\sinh (\gamma _{n}(2\pi -t_{1}))}
{\gamma _{n}(\cosh (2\pi \gamma _{n})-\cos (2\pi \beta _{n}))},
\end{equation}
 where
\begin{equation}
t_{1}=\frac{2{\pi }R_{1}}{L},\, \, t_{2}=\frac{2{\pi }}{L}(R_{1}
\cos \theta +R_{2}),\, \, \beta _{n}=n\cos \theta ,\, \, 
\gamma _{n}=\sin \theta \sqrt{n^{2}+L^2/(2\pi \lambda)^2}.
\end{equation}
This expression is simpler than the one used by Nordborg and Blatter
since it does not have different expressions for even and odd n. It
also converges faster in certain regions.

\section{Permutation sampling}

Essentially the same method is used for permutation sampling as was
used in Ref. \onlinecite{nordborg}. The only difference is that we use 
permutation
space of only the neighboring lines. This is so because even if we
were able to get a permutation step, involving a large number of lines,
accepted at the first stage of the algorithm, it would be very likely
to get rejected at the following stages. So we work with only 3-5
lines. Sufficiently long segments (typically 5 planes) of a number
of lines were cut. Care should be taken to make sure that chosen FLs
are the nearest neighbors in the plane where the reconnection of FLs
is going to take place. These points can be implemented easily with
the concept of linked lists and pointers \cite{tildesley}. Also,
care is to be taken that even though a FL may be far away from
some other FL in the rhombically shaped unit cell, it can still permute
with it through one of the images of the latter. These few simple points
are very important to implement the whole procedure correctly. Just for
a check, we tried with the sampling procedure given in Ref. 
\onlinecite{ceperley}.
This gave results in good agreement with the sampling procedure given
above.

\end{document}